\magnification=1100
\vsize=22truecm
\parindent=1.2truecm
\baselineskip24truept
\overfullrule0pt 
\def\cl{\centerline}
\def\nt{\noindent}
\def\bs{\vskip.3in}
\def\ms{\vskip12pt}
\def\ss{\vskip6pt}

\font\med=cmbx10 at 12pt

\cl{\med {THE dS/CFT CORRESPONDENCE}}
\ss

\cl{\med {AND THE BIG SMASH}}
\ss

\ms
\cl{\med{by}}

\ss

\cl{\med {Brett McInnes}}
\baselineskip14truept
\cl{\bf Department of Mathematics}

\cl{\bf National University of Singapore}

\cl{\bf 10 Kent Ridge Crescent}

\cl{\bf Singapore 119260}

\cl{\bf Republic of Singapore}

\cl{matmcinn@nus.edu.sg}

\bs
\baselineskip15truept
\nt {\bf Abstract.}\ \ Recent observations suggest that the cosmological equation-of-state parameter $w$ is close to $-1$. To say this is to imply that $w$ could be slightly {\it less} than $-1$, which leads to R.Caldwell's ``Phantom cosmologies". These often have the property that they end in a ``Big Smash", a final singularity in which the Universe is destroyed in a finite proper time by excessive {\it expansion}. We show that, classically, this fate is not inevitable: there exist Smash-free Phantom cosmologies, obtained by a suitable perturbation of the deSitter equation of state, in which the spacetime is in fact asymptotically deSitter. [ Contrary to popular belief, such cosmologies, which violate the Dominant Energy Condition, do 
{\it not} necessarily violate causality.] We also argue, however, that the physical interpretation of these classically acceptable spacetimes is radically altered by ``holography'', as manifested in the dS/CFT correspondence. It is shown that, if the boundary CFTs have conventional properties, then recent ideas on ``time as an inverse renormalization group flow" can be used to rule out these cosmologies. Very recently, however, it has been argued that the CFTs in dS/CFT are of a radically unconventional form, and this opens up the possibility that Smash-free Phantom spacetimes offer a simple model of a ``bouncing" cosmology in which the quantum-mechanical entanglement of the field theories in the infinite past and future plays an essential role.

\ms
\nt{\bf 1.\ \ INTRODUCTION}
\ms
The rapidly accumulating evidence in favour of cosmic acceleration [see, for example, [1]] has shown that much of the energy in the Universe violates the Strong Energy Condition [SEC]. In view of this, it is natural to question [2] the status of {\it all} of the energy conditions: if the failure of the SEC was foreseen by few, surely it is unwise to refuse to contemplate the loss of its weaker counterparts. If they too are violated, then we can begin the search for still weaker but useful replacements. If they are {\it not} violated, then the question becomes: why not? If the SEC is so spectacularly wrong, then we need theoretical reasons for continuing to believe in the other conditions.
\ss
The main surviving energy condition is the Dominant Energy Condition [DEC], which requires the cosmological density and pressure to satisfy
$$ \rho \ge |p|. \eqno{(1)}$$
The observational status of this condition is by no means secure, as Caldwell [3] has pointed out. The cosmological data are usually expressed in terms of the parameter $w$, defined by
$$ p = w\rho; \eqno{(2)}$$
of course, $w = -1$ for ``cosmological constant dark energy" $-$ let us call it ``$\Lambda$ dark
energy" to distinguish it from other forms of dark energy such as quintessence. The current data are summarized in Figure 4 of [4]. They are consistent with $w = -1$, {\it but} they are certainly also consistent with lower values, which would, if confirmed, violate the DEC. [Cosmological models with $ w < -1$ are known as ``Phantom cosmologies" [3].]
\ss
Theoretical efforts to come to terms with this possibility have largely focused on scalar fields, particularly non-minimally coupled scalars [2][5][6][7]. Particularly striking, because it is founded on string theory, is the DEC-violating cosmology of Kachru and McAllister [8], who study brane cosmology in the context of type IIB theory. The effective action for an observer on the brane is that of a scalar-tensor theory with negative $\rho + p$. Thus, it seems that string theory can accommodate violations of the {\it four-dimensional} DEC: the understanding is presumably that the full, higher-dimensional theory in some way regulates the various kinds of misbehaviour traditionally associated with such violations. We are therefore obliged to reconsider those ``traditional" objections.
\ss
There are several such objections, stated for example in [9]. The first and most potent is the idea that $w < -1$ leads to violations of causality. The easiest way to see why this might be of concern is to recall that the local speed of sound in a perfect fluid is given by $[{dp\over d\rho}]^{1/2}$ times the speed of light. Thus, fluids in which pressure dominates density are in danger of having a speed of sound which exceeds that of light. However, note that the {\it cosmological} equation of state only applies to strictly isotropic situations---that is, $p$ and $\rho$ in equation (2) are not allowed to depend on spatial position, only on time. Hence (2) is not a full ``equation of state", and nothing can be deduced from (2) {\it alone} about the speed of sound. A more subtle point is that this argument assumes that the ``fluid" is capable of transmitting signals in the first place. This is not the case for $\Lambda$ dark energy, since its density and pressure are absolutely constant in space and time $-$ it is clearly nonsensical to speak of an ``energy flow" in $\Lambda$ dark energy, or to allow the latter to contribute to the energy flow of any other fluid which may be present. In computing the speed of propagation of signals in such a fluid, one must therefore exclude the contributions of $\Lambda$ dark energy to $\rho$ and $p$. In short, values of $|w|$ above unity {\it may} lead to causality problems $-$ but they need not.
\ss
A more sophisticated version of the causality objection is based on the observation that a theorem of Hawking and Ellis [pages 91-94 of [10]; see also [11]] states that if a stress-energy-momentum tensor $T$ satisfies the DEC and vanishes on a closed achronal set $S$, then $T$ must also vanish on the domain of dependence of $S$. Roughly speaking, this means that the DEC forbids anything to enter the region of spacetime controlled causally by $S$. The answer to this is simple: the theorem does not have a converse. There are examples of spacetimes which satisfy the conclusion of the theorem without satisfying its assumption. The best known of these is the now very familiar Anti-deSitter space, which, with its negative density, strongly violates the DEC; but no-one would claim that AdS ``violates causality". The fact that signals can reach a bulk observer from infinity does not change this fact: such signals cannot enter the domain of dependence of $S$ unless they register on $S$ itself. Thus, while the Hawking-Ellis theorem opens up the {\it possibility} of causal misbehaviour in Phantom spacetimes, it does not {\it require} it. Such spacetimes must be checked on a case-by-case basis, by drawing the Penrose diagram. Indeed, the Penrose diagram of Anti-deSitter space shows that while this spacetime has several strange properties, causal misbehaviour is not among them. [It is sometimes suggested that the DEC should be modified so as not to exclude Anti-deSitter space, usually by requiring it to apply to null but not timelike vectors. The physical motivation for this is, to say the least, problematic $-$ should we be more interested in the findings of ``observers" with null worldlines than in those of real observers? Null vectors are relevant to the energy conditions only through continuity arguments relating to timelike vectors. In any case, there is no reason to believe that the Hawking-Ellis theorem continues to hold when this weaker condition is used, and hence there is no reason to expect causality problems if this version of the DEC is violated.]
\ss
A completely different objection to spacetimes which violate the DEC is that they frequently lead to cosmological singularities of a peculiarly unpleasant kind. Caldwell [3] [see also [12]]
constructs a Phantom cosmology which is {\it not} asymptotically deSitter, no matter how close $w$ may be to $-1$. If we assume that Phantom energy comes to dominate after a matter-dominated period ends at $t = t_m$, then in the model of [3] we have 

$$ a(t) = a(t_m)[-w + (1+w)({t \over t_m})]^{2 \over 3(1+w)}, \eqno{(3)}$$

\nt where $a(t)$ is the usual FRW scale factor, where $w$ is assumed to be {\it constant}, and where the spatial sections are assumed to be flat. We note that this diverges at the finite proper time 
$$ t = ({w \over 1+w})t_m, \eqno{(4)}$$
so this universe is not asymptotically deSitter; it comes to an end. But this is not a ``Big Crunch" $-$ the Universe is destroyed not by excessive contraction but rather by excessive {\it expansion}.  Nevertheless, as the universe evolves towards this strange disaster, its density must {\it increase}, since this is a characteristic property of all Phantom cosmologies, as can be seen from the ``conservation" [10] equation:
$${\mathop{\rho}\limits^\cdot} + 3(\rho + p){\mathop{a}\limits^\cdot}/a = 0. \eqno{(5)}$$
Indeed, in the model of [3], the constancy of $w$ allows equation (5) to be integrated to yield
$$ \rho = K[-w + (1+w)({t \over t_m})]^{-2}, \eqno{(6)}$$
where $K$ is a positive constant, so $\rho$ too diverges as the time given by (4) is approached. Thus the final state is singular, with ``infinite" density and pressure, despite the fact that the universe expands more and more rapidly. Instead of suffering a ``crunch", the universe is ``smashed", yet this smashing is no less singular than a crunch. Let us speak of a ``Big Smash" cosmology.
\ss
It is evidently a matter of urgent cosmic concern to determine whether the Big Smash is a generic aspect of Phantom cosmology. It is usually felt that classical singularities can be tolerated only to the extent that one can hold out hope of their being resolved quantum-mechanically, when the universe is very small. The Big Smash, however, seems to offer no such hope (unless perhaps one could treat it, in some unknown way, as a ``T-dual" configuration to a Big Crunch). Thus it seems that by violating the DEC we are finally losing all hope of eliminating singularities from cosmology. This may be a strong objection, but we notice that Caldwell's model {\it assumes} that $w$ is constant. We can therefore hope to answer this objection simply by allowing $w$ to vary with time, a frequently discussed possibility which arises, for example, in quintessence models. 
\ss
To summarize: the widespread unease with DEC-violating cosmologies is based on strong physical arguments $-$ all of which, however, have loopholes. The string-based Kachru-McAllister Phantom cosmology prompts the question: are there Phantom spacetimes which avoid all of the traditional objections? We shall see that the answer is positive. There are simple DEC-violating spacetimes with non-constant $w$ in which all signals propagate causally, which have perfectly well-behaved Penrose compactifications, and which do not have a Big Smash. This is of considerable interest in itself.
\ss
The [classically] well-behaved Phantom cosmology constructed here is asymptotically deSitter in a very explicit sense. As is well known, much effort has been devoted to attempts to gain a ``holographic" understanding of such spacetimes by constructing a ``dS/CFT correspondence" [13][14][15]; see [16] for the early references, [17] for a discussion of the aspects that are relevant here. Having established that our cosmology evades the traditional objections to violations of the DEC, we next ask: what is the nature of ``holographic Phantom cosmology?
\ss
We argue that holography drastically changes the physical interpretation of Smash-free Phantom cosmologies, and at first sight it seems that the change is in the direction of decreased physical plausibility. The argument is conceptually simple. As was emphasised in [18] and [19], the Euclidean formulation of holographic theories {\it cannot easily accommodate boundaries which are topologically disconnected}. This is a basic principle of holography, since a disconnected boundary inhabited by two or more completely independent field theories would undermine the very idea that the {\it connected} bulk gives a complete account of the boundary. To see this, imagine varying the parameters of one boundary theory but not those of the others. Does the bulk theory change? A disconnected boundary {\it may} be tolerable in a Lorentzian theory, since there may be a causal relationship between the boundary components (or since an event horizon may effectively disconnect the bulk, as in the asymptotically Anti-deSitter version of the Kruskal-Szekeres spacetime [20][21]),  but this does not apply to the Euclidean version. In fact, there are powerful theorems [19][22][23] which, under precise conditions which must be carefully verified, actually {\it forbid} the boundary to be disconnected in the Euclidean case, {\it unless}, of course, the bulk is also disconnected. 
\ss
Subsequent developments, particularly Maldacena's [21] analysis (see also [20]) of the holography of black holes, have shown that there can perhaps be exceptions to this rule in very specific circumstances, namely if the independent boundary theories are ``entangled" quantum-mechanically, since the entangled state is prepared using a Hartle-Hawking construction which may disconnect the boundary of the Euclidean part of the space.  The Smash-free Phantom cosmologies actually have Euclidean versions which violate precisely one condition of the relevant theorem, and they all have Euclidean versions with disconnected boundaries. They are therefore directly contradictory to holography, {\it unless} the bulk can in some way be regarded as an effectively disconnected space $-$ or unless some
kind of quantum-mechanical entanglement allows the same kind of ``exception to the rule" as in the case considered by Maldacena. This last possibility has recently come to seem much more plausible, and it may well play an essential role in ``holographic Phantom cosmology".
\ss
Section 2 presents the Smash-free Phantom cosmologies. They are deduced, in a very straightforward way, from a physically motivated modification of the equation of state. Section 3 gives a brief analysis of the circumstances under which disconnected holographic boundaries might and might not be tolerable. We discuss the cosmologies presented in Section 2 in this light. Section 4 proposes an analysis of this situation using the speculations of Strominger [24] and of Balasubramanian et al [25] on ``time as an inverse renormalization group flow". We propose that ``time flows the wrong way" in an expanding Smash-free Phantom cosmology, and that this effectively disconnects the bulk. This resolves the conflict with holography, at the cost of producing a cosmology in which the Universe necessarily contracts. The alternative, preferable, but still speculative interpretation in terms of an entanglement between the infinite past and the infinite future is also briefly considered. 
\bs
\nt{\bf 2.\ \ SMASH-FREE PHANTOM COSMOLOGY}
\ms
\ss
We have seen that the simplest Phantom cosmologies destroy the Universe in a singularity which is even more extravagant and bizarre than the Big Crunch, namely in a Big Smash. We shall now show how to construct Phantom cosmologies with no Big Smash. These cosmologies are of interest not merely because of the observational data (though these [4] are certainly very suggestive) but also because the non-singular ``bouncing" cosmologies which have recently attracted much attention (see [26] and its references) necessarily violate the DEC during (at least) some part of their history. The spacetimes we shall consider can be expected to play the same role in these ``bouncing" cosmologies as deSitter space plays in inflation. They are in fact asymptotically deSitter, in the sense that they are indistinguishable from deSitter space at very late [and very early] times. We shall think of them as being generated by the introduction of some kind of  (necessarily rather exotic) matter into a background with a positive cosmological constant. Ideally, one would analyse this exotic matter in terms of string or M theory, but, as is notorious, we do not fully understand how to obtain deSitter space itself from those theories; still less do we know what manner of perturbations, to be interpreted cosmologically as exotic matter, those theories can generate around a deSitter-like background. In [3], Phantom energy is represented by reversing the sign of the kinetic term in a quintessence action. There are hints that this might indeed be motivated by string theory, as for example in Hull's [13] approach to embedding deSitter space in M-theory (where kinetic terms with the ``wrong" sign arise), but there are obvious difficulties with such an approach. Fortunately it is possible to proceed on the basis of more general assumptions.
\ss
The ``conservation" equation (5) may be written as
$$ {d\rho \over da} = -3(\rho + p)/a, \eqno{(7)}$$
and so we see that, in every Phantom cosmology, $\rho$ must increase with $a(t)$. Thus, the only way to avoid a Big Smash (and obtain an asymptotically deSitter geometry) is to arrange for $\rho$ to approach $3/8\pi L^2$, its asymptotic deSitter value (where $1/L^2$ is the constant positive sectional curvature of deSitter space), from below; we must have 
$$ \rho = {3\over 8\pi L^2} - S(a), \eqno{(8)}$$
where $S(a)$ is a positive function which decreases monotonically to zero as $a(t)$ tends to infinity. Physically, equation (8) means that asymptotically deSitter Phantom cosmologies are necessarily obtained by introducing ``matter" with a negative energy density into a deSitter-like background. In view of the uncertainties, we shall not try to specify the nature of this ``matter"; indeed, it may not be matter in the usual sense, but rather [for example] an effective representation of higher-order terms resulting from the inclusion of more complicated curvature terms in the Lagrangian. [That such terms can lead to effective violations of the DEC has been emphasised and applied in [26].] We shall simply assume that this ``matter", $\psi,$ satisfies equation (2) with a negative $\rho_{\psi}$ and a positive $p_{\psi},$ and thus a negative $w_{\psi}$. For simplicity, we shall take $w_{\psi}$ to be constant.  The total pressure [deSitter plus that of $\psi$] is now
$$ p = p_{dS} + p_{\psi}$$
$$ = -\rho_{dS} + w_{\psi}\rho_{\psi}$$
$$ = w_{\psi}\rho - {3(1+w_{\psi})\over 8\pi L^2}, \eqno{(9)}$$
where $\rho$, given by (8), is the total density. Thus equation (9) is the appropriate form of the equation of state in asymptotically deSitter Phantom cosmology. It can be written as in equation (2), with $w < -1$, but then $w$ will {\it not} be a constant. This is crucial: it is only with a non-constant $w$ that a Big Smash can be averted. The key point here is that a superposition of two fluids, each with constant but different $w$, does not itself have constant $w$.
\ss
Now since $\rho + p < 0$ in Phantom cosmology, and since $\rho + p = \rho_{\psi} + p_{\psi},$
it follows that $\rho_{\psi} + p_{\psi} = \rho_{\psi}(1 + w_\psi)$ has to be {\it negative}; hence, in view of the fact that $\rho_{\psi}$ is negative and $p_{\psi}$ is positive, we must have $-1 < w_{\psi} < 0$, as with ordinary quintessence [for example]. Notice that this inequality means that $\psi$ itself has a pressure which does not dominate its density [that is, the pressure never exceeds $|\rho|$]; hence there is no reason to expect that its sound speed will exceed the speed of light. [Again, to prove this one would have to set up a detailed model of $\psi$, with a genuine equation of state that would allow a computation of the speed of sound: the point is that there is no reason whatever to expect that this speed will necessarily be superluminal.] Since $\Lambda$ dark energy has strictly constant density and pressure, and therefore cannot influence the speed of signal propagation, we see that the combined fluid does not violate causality. 
\ss
In order to keep track of the signs, and for later convenience, we introduce a constant parameter $\gamma$ defined by
$$ \gamma = 3(1 + w_{\psi}). \eqno{(10)}$$
The bounds on $w_{\psi}$ are then expressed by 
$$ 0 < \gamma < 3, \eqno{(11)}$$
and the equation of state now reads
$$p = -\left(1 - {\gamma \over 3}\right) \rho - {\gamma \over {8\pi L^2}}. \eqno{(12)}$$
Substituting (8) into (12), we obtain
$$p = - {3\over 8\pi L^2} + \left(1 - {\gamma \over 3}\right) S(a), \eqno{(13)}$$
and thus
$$\rho + p = - {\gamma \over 3} S(a). \eqno{(14)}$$
Since, as we shall soon find, the maximum possible value of $S(a)$ (for all $\gamma$) is $3/8\pi L^2$, it follows that the minimum value of $\rho + p$ is $-\gamma /8\pi L^2$, and so we see that the parameter $\gamma$ measures the extent to which the DEC is violated. We shall therefore normally think of $\gamma$ as a small number, since we do not wish to violate the DEC too flagrantly.
\ss
We claim that (12), unlike (2), can lead to an asymptotically deSitter FRW metric. We shall assume that the spatial sections are exactly flat and compact. Apart from the well-known observational evidence in favour of flatness [1], there are theoretical motivations connected with holography, as we shall discuss later. These considerations also motivate the requirement of compactness, though of course non-trivial spatial topology is also of observational interest. To be specific, we shall take the spatial sections to be cubic tori of minimum circumference $2\pi A$, where $A$ is a constant. Taking $a(t)$ to be the scale factor, and parametrising the torus by angles, we therefore have a metric of the general form
$$-dt \otimes dt + a(t)^2A^2(d\theta_1 \otimes d\theta_1 + d\theta_2 \otimes d\theta_2 + d\theta_3 \otimes d\theta_3). \eqno{(15)}$$
Substituting (14) into the ``conservation" equation (7), we obtain 
$${dS\over da} = - \gamma S/a, \eqno{(16)}$$
with solution
$$S = \alpha \ a^{-\gamma}, \eqno{(17)}$$
where $\alpha$ is a positive constant. Now if we are to obtain an asymptotically deSitter cosmology, that is, one which contracts in the past and expands in the future, there must be a time $-$ let us take it, as in deSitter space itself, to be $t = 0$ $-$ when the time derivative of $a(t)$ is momentarily zero. The Einstein equation for FRW models with {\it flat} spatial sections is simply
$$3{\dot a}^2/a^2 = 8\pi \rho , \eqno{(18)}$$
and so we see that $\rho$ must be zero at $t = 0$. Comparing (17) with (8) and noting that (by definition of $A$) we must have $a(0) = 1$, we see that $\alpha$ is now fixed; so we have
$$\rho = {3\over 8\pi L^2} - {3\over 8\pi L^2} a^{-\gamma}, \eqno{(19)}$$
and 
$$p = - {3\over 8\pi L^2} - \left( {\gamma \over 3} - 1\right) {3\over 8\pi L^2} a^{-\gamma}. \eqno{(20)}$$
Substituting (19) into (18) we find that there is an elementary solution:
$$a = {\cosh}^{\left( {2\over \gamma}\right)} \left({\gamma t \over 2L}\right), 
\eqno{(21)}$$
so we have a family of spacetimes parametrised by $\gamma$ and $A$:
$$g(\gamma ,A) = -dt \otimes dt + A^2 {\cosh}^{\left({4\over \gamma}\right)}
	\left( {\gamma t \over 2L}\right) (d\theta_1 \otimes d\theta_1 + 
d\theta_2 \otimes d\theta_2 + d\theta_3 \otimes d\theta_3). \eqno{(22)}$$
We shall refer to this spacetime as $Ph(\gamma, A)$, the Phantom spacetime with parameters $\gamma$ and $A$. {\it We regard these spaces as the Phantom analogues of deSitter space,} in the sense that these are the simplest possible non-singular Phantom metrics. All of the $Ph(\gamma, A)$ ``tend to deSitter space", in its flat slicing, in the following precise sense: for all $\gamma$, we find that if we put $x = 2^{-({2\over \gamma})}A\theta_1$, $y = 2^{-({2\over \gamma})}A\theta_2$, $z = 2^{-({2\over \gamma})}A\theta_3$, then for large $t$ the metrics all tend to
$$g(dS_4; \hbox{future}) = -dt \otimes dt + e^{2t\over L} (dx \otimes dx + dy \otimes dy + dz \otimes dz), \eqno{(23)}$$
that is, the metric of the ``future triangular half" of deSitter space, in the flat slicing. Note carefully that we obtain this precise agreement because we assumed that the spatial sections of our Phantom cosmology should be flat. (The fact that $x$, $y$, and $z$ have finite ranges is observationally irrelevant at very late times, because eventually a cosmological horizon will conceal the periodic identifications; see below.)
\ss
To see that the metrics (22) are all distinct from each other (and from deSitter space), we compute the scalar curvature: it is
$$R(g(\gamma , A)) = {12 \over L^2} - {3\over L^2} (4 - \gamma) {\rm sech}^2 
	\left( {\gamma t \over 2L} \right); \eqno{(24)}$$
note that $12/L^2$ is the scalar curvature of deSitter space, the asymptotic value. The ratio of the minimum scalar curvature to the scalar curvature of deSitter space is then
$$R(g(\gamma ,A))_{\rm min}/R(dS_4) = \gamma /4, \eqno{(25)}$$
which reveals the geometric significance of $\gamma$, and which shows that all of our solutions are indeed distinct from each other and from deSitter space.
\ss 
Returning to (22), note that the graph of ${\cosh}^{\left({2\over \gamma}\right)}
	\left( {\gamma t \over 2L}\right)$ has the same general shape as that of $\cosh (t/L)$, so the qualitative features of our cosmologies resemble those of deSitter space, in its full global form:
$$g(dS_4) = -dt \otimes dt + L^2 {\cosh}^2 \left( {t\over L}\right) 
\left[ d\chi \otimes d\chi + {\sin}^2 (\chi)(d\theta \otimes d\theta + {\sin}^2(\theta) d\phi \otimes d\phi)\right]. \eqno{(26)}$$
That is, it contracts from ``infinity" in the past to a minimum size , and then expands out again to an ``infinite" future. There are two principal differences, however. First, in the deSitter case, the minimum size of the Universe is determined by $L$, that is, by the deSitter radius; in our case, by contrast, the minimum size is determined by $A$, and so can be prescribed independently of $L$. Second, if $\gamma$ is small, then the graph of ${\cosh}^{\left({2\over \gamma}\right)}
	\left( {\gamma t \over 2L}\right)$ is much flatter than that of $\cosh (t/L)$, so the Universe contracts to its minimum size, and expands away from it, much more slowly than in the deSitter case; the deSitter rate of acceleration, $1/L^2$, only takes over at a very late (or very early) proper time, as one sees from the Raychaudhuri equation:
$${\hbox{\"a}}/a = {1\over L^2} + {1\over L^2} \left( {\gamma \over 2} - 1\right) a^{-\gamma}.\eqno{(27)}$$
The Penrose diagram is therefore a rectangle with width determined by $A$ and height determined by the full extent of conformal time,
$$\int^\infty_{-\infty} {dt \over {\cosh}^{\left({2\over \gamma}\right)} \left({\gamma t \over 2L}\right)} = {2L \over \gamma} \int^\pi_0 {\sin}^{\left({2\over \gamma}-1\right)}  (\psi) d\psi , \eqno{(28)}$$
where the variable has been changed to make it clear that the integral converges. This quantity actually {\it increases} as $\gamma$ becomes smaller. For any given value of $A$, we can, by making $\gamma$ small enough, produce a Penrose diagram which is ``tall and thin". Again, it is perfectly clear that there are no causality violations in this spacetime.
\ss
Although it is not our objective here to construct a fully realistic cosmological model $-$ to do that, one would of course have to add more conventional matter to the spacetime $-$ it is worth noticing that $Ph(\gamma, A)$ actually has some interesting and agreeable properties. By taking $A$ to be small, one can produce a Universe which in the past was as small as one might wish, but non-singular ; but then by taking $\gamma$ sufficiently small, so that the Penrose diagram is ``tall and thin", one can do away with particle horizons and so solve the horizon problem. (As is well known, a single inertial observer can exchange information with only half of the inertial observers in deSitter space (corresponding to the ``northern (or southern) diamond" in the conformal diagram); but here such an observer can interact with the entire Universe.) Now of course this kind of advantage arises in many cosmologies with flat, compact spatial sections, but the problem is that, in the usual non-Phantom cosmologies, such non-trivial spatial topology has a definite signature, which has not been observed in the flat case [27]. However, the novelty here is that, as in any asymptotically deSitter cosmology, a future horizon will eventually appear as the cosmic acceleration increases, so ultimately the edge of the fundamental domain will be swept away beyond the horizon; the direct observational evidence for the fact that space is finite is thus concealed; the spacetime is, ultimately, observationally indistinguishable from the future triangular half of deSitter space, with metric given by equation (23). One can have one's topologically non-trivial cake and eat it. 
\ss
To summarize: using the equation of state (9) or (12), we have found  very simple Phantom cosmologies with no causality problems and no Big Smash. The space-time geometry resembles a version of global deSitter space (but with flat compact spatial sections) which has been stretched in the time direction and narrowed in the spatial directions. The Universe contracts from ``infinity" very much like deSitter space, but as it becomes small it ``goes into hibernation", solving the horizon problem, as it slowly evolves through its minimum size; eventually the expansion picks up pace, and at late times the structure is indistinguishable from that of deSitter space. As with deSitter space, the conformal boundary consists of two disconnected pieces $-$ two tori instead of two spheres. That is in fact a major reason for our interest in these spacetimes: they allow us to investigate concrete, exact examples of connected cosmological ``bulks" with two boundaries. At this point, however, we conclude that we have found Phantom cosmological models which are at least as acceptable as deSitter space; or so it seems.
\bs
\nt{\bf 3.\ \ CONNECTEDNESS OF THE BOUNDARY IN HOLOGRAPHY}
\ms
The basic idea of holography, as embodied in the AdS/CFT correspondence, is that a bulk gravitational theory is entirely equivalent to a boundary (conformal) field theory. An immediate fundamental objection to this idea, raised already in [18], is that a {\it connected}, compact manifold-with-boundary can easily have a {\it disconnected} boundary. This obviously militates against any kind of one-to-one correspondence; worse, it is not entirely clear how holography can be internally consistent in such cases, since the boundary theories (defined on timelike conformal boundaries) are typically decoupled from one point of view, yet coupled from the other.
\ss
A possible resolution of this problem was proposed in [18] and confirmed in [28], where it was shown that, under certain conditions, disconnected boundary components of an asymptotically AdS spacetime are necessarily concealed from each other by an event horizon. Thus, the boundary theories are decoupled from both points of view. A useful way of putting this is to say that, while the bulk in this case is connected in the usual topological sense, it is {\it effectively disconnected} from a physical point of view. The deep significance of this issue is illustrated by the analyses of [20] and [21], where black hole singularities are studied by means of independent yet {\it entangled} conformal field theories on a disconnected boundary. In accordance with [28], the bulk is effectively disconnected by an event horizon, so the conformal field theories are indeed independent; but their entanglement is due to the fact that they are defined on different asymptotic regions of the {\it same} black hole singularity. The distinction between topological connectedness and effective disconnectedness clearly lies at the heart of this situation.
\ss
When we turn to the dS/CFT correspondence, similar issues arise, sharpened by the fact that (unlike Anti-deSitter space) the deSitter spacetime already has a disconnected conformal boundary [29]. The resolution here is different, but in the same spirit: given a point on the sphere at past infinity, there is a null curve connecting it to the antipodal point at future infinity. The past-future correlators are then non-trivial [14],[15]; the two boundary components are far from being independent. Instead of finding that the bulk is effectively disconnected, we find that the boundary is {\it effectively connected} $-$ and again the one-to-one bulk/boundary correspondence is maintained. 
\ss
To summarize thus far: topologically disconnected conformal boundaries are, in the Lorentzian case, compatible with holography {\it provided} that either the bulk is effectively disconnected or the boundary is effectively connected. A very deep illustration of this philosophy is provided by an example which at first sight appears to contradict it, namely the Karch-Randall [30][31] theory. This is a brane-world theory, in which an $AdS_4$ brane is embedded in $AdS_5$ through the foliation of the latter by the former: the $AdS_5$ metric may be written unconventionally as
$$g^{AdS_5} = dz \otimes dz + {\cosh}^2 \left({z \over L}\right) g^{AdS_4}. \eqno{(29)}$$
The brane, at $z = z_0$, cuts off $AdS_5$, and so the bulk (the remaining part of $AdS_5$) {\it seems} to have ``two boundaries" [21][32], namely $B_1$, the brane itself, and $B_2$, the portion of the $AdS_5$ boundary that has not been discarded. However, the full boundary of this bulk also includes the boundary of $B_1$, which is also the boundary of $B_2$. (This corresponds to the large dots in Figure 6 of [30].) This extra piece plays a crucial role in the physics: for AdS/CFT tells us how $B_2$ accounts for the bulk, and it should also tell us how the brane boundary accounts for the brane physics; the fact that the brane boundary is also the boundary of $B_2$ must be related to the Duff-Liu complementarity [33][32]. At any rate, when we assemble all three components of the boundary, we find that it is connected; this is clear from Figure 6 of [30]. This is in agreement with the above arguments, since $B_1$ and $B_2$ can communicate through the bulk. (Notice that equation (29) does give the impression that the boundary of $AdS_5$ is disconnected, with one component at $z = \infty$ and the other at $z = -\infty$, but this is incorrect, of course; we have to bear in mind that, in this case, the slices are themselves infinitely large and reach out to the boundary.)
\ss
Euclidean methods have been extremely successful [34] in the study of AdS/CFT, and it is natural to ask how the above arguments translate into this language. The absence of causality eliminates the distinction between actual and effective connectedness, and so the arguments against disconnected boundaries become all the more pointed. Exceptions must be well justified.
\ss
In [21], as was mentioned above, an asymptotically AdS Kruskal-Szekeres spacetime is described holographically by two identical but non-interacting CFTs on the two components of the conformal boundary, in a particular entangled state which is prepared using the well-known Hartle-Hawking construction. That is, the Lorentzian and Euclidean versions of the manifold are each ``halved" and then one Lorentzian half is joined to the ``top" of a Euclidean half. The full Euclidean black hole space has a connected boundary, but the half used here does not. However, the very fact that we are preparing an entangled state means that the arguments against disconnected boundaries are very much less persuasive here. While the correct procedure is not yet fully clear, it is reasonable to adopt the following working rule, in an effort to reconcile the concerns of [19] and [21]: in applying Euclidean methods to AdS/CFT and dS/CFT, we should reject connected spaces with topologically disconnected boundaries {\it unless we can produce evidence suggesting that the CFTs on the various boundary components are entangled.} With this principle in mind,
let us proceed to consider the Euclidean versions of the Smash-free Phantom cosmologies.
\ss
In Euclideanizing asymptotically deSitter spacetimes, we have to be aware that they differ very significantly from asymptotically Anti-deSitter spacetimes. Traditionally, Euclideanization is performed by ``complexifying" time. Of course, this greatly modifies the geometry in the time direction, while leaving the spatial directions relatively unscathed $-$ and this is undoubtedly one reason why Euclidean methods work so well in AdS/CFT. For the boundary of AdS is at {\it spatial} infinity, so complexifying {\it time} does not disturb the structure of the spacetime too drastically from the holographic point of view. Since, on the other hand, the deSitter boundary is at temporal infinity, it is clear that the traditional Euclideanization procedure will not be fully adequate for asymptotically deSitter spaces. We need to Euclideanize in a way which changes space instead of time. That is, instead of mapping the signature from ($ \ - \ + \ + \ +$) to 
($ \ + \ + \ + \ +$), we map it to ($ \ - \ - \ - \ -$), leaving time unchanged and ``complexifying" space [35]. This may seem odd, but it is in fact no more than a simple extension of a technique familiar to students of space-time ``bubbles of nothing". [See for example [36] and references therein.] The standard technique, introduced long ago by Witten, is ``double analytic continuation", which simply involves complexifying {\it two} coordinates instead of one. Here we wish to use a ``triple analytic continuation". Of course, physics can no more favour ($ \ + \ + \ + \ +$) signature over ($ \ - \ - \ - \ -$) than it can favour the relativist's ($ \ - \ + \ + \ +$) over the particle theorist's ($ \ + \ - \ - \ -$), though one must change signs suitably in various places. Naturally, if we wish to go beyond using ``multiple analytic continuation" as a mere solution-generating technique, and to use it in a path-integral formalism, then we must specify how to take care of the inevitable $\pm$ signs and factors of $i$. Here in fact triple analytic continuation works better than the more familiar double continuation, simply because $(\pm i)^3 = \mp i$; thus, just as a single analytic continuation causes the action to pick up a factor of $\pm i$, so here the action picks up a factor of $\mp i$, the sign being chosen so that the Euclidean path integral should converge. On the other hand, in five or other odd spacetime dimensions this method works no better than double analytic continuation. Notice however that five-dimensional spacetimes are usually taken to be static: brane-worlds, for example, typically reside in $AdS_5$. Since triple analytic continuation is specifically designed for dynamic spacetimes (with the intention of preserving the structure of the time evolution), this should not be a problem in practice.
\ss
It is of course a mere matter of convention that Riemannian geometry employs positive rather than negative definite metrics; though we must remember to take the absolute value when measuring lengths, and that a ``negative sphere" has negative sectional curvature, while a ``negative hyperbolic space" has {\it positive} sectional curvature. Thus, the negative hyperbolic metric with signature ($ \ - \ - \ - $) and sectional curvature $+1/L^2$ is
$$g^{-H^3}_{\left(+ {1\over L^2}\right)} = - L^2d\chi \otimes d\chi - L^2 {\sinh}^2(\chi)(d\theta \otimes d\theta + {\sin}^2(\theta)d\phi \otimes d\phi). 
\eqno{(30)}$$
Now if we take $g(dS_4)$, given by equation (26), and complexify $\chi$ instead of $t$, we get 
$$-dt \otimes dt - L^2 {\cosh}^2 \left( {t\over L}\right) 
\left[ d\chi \otimes d\chi + {\sinh}^2 (\chi)(d\theta \otimes d\theta + {\sin}^2(\theta) d\phi \otimes d\phi)\right], \eqno{(31)}$$
which is precisely
$$-dt \otimes dt + {\cosh}^2 \left({t \over L}\right)g^{-H^3}_{\left(+ {1\over L^2}\right)}.
\eqno{(32)}$$
Thus the negative Euclideanization of deSitter space is obtained simply by replacing the three-spheres of sectional curvature $+1/L^2$ by negative hyperbolic spaces, {\it also} of sectional curvature $+1/L^2$. As we intended, the characteristic deSitter ``time geometry" (large in the past, contracting to a minimum size, and then expanding out to infinity) has been preserved by negative Euclideanization. If we had complexified time, the result would have been a four-sphere, with no boundary at all: the temporal infinities would have been extirpated. That is desirable for some purposes [37], but not if we want to study a bulk-boundary duality.
\ss
Now (32) {\it seems} to represent a negative Euclidean space with a conformal boundary in two disconnected pieces, one at $t = -\infty$, and the other at $t = +\infty$. But this is where we must remember the fact that negative Euclideanization modifies the structure of the spatial directions rather than the temporal direction. For now the ``spatial" directions in (32) are infinitely large, and they reach out to the boundary. This has the crucial consequence that the two boundary components bend around and touch. In fact, (32) is just the metric of $-H^4$, foliated by negative hyperbolic subspaces, $-H^3$. That is, $-H^4$ is the negative Euclideanization of $dS_4$. (This is in agreement with the fact that both $dS_4$ and $-H^4$ are spaces of positive curvature, $+1/L^2$. Compare with equation (29), and recall that the {\it positive} Euclideanization of $AdS_n$ is $+H^n$.) Picturing $-H^4$ (which has $-S^3$ as its conformal boundary) as a disc, this foliation can be visualised by imagining ``phases of the moon". The slices all intersect at infinity along that equator of the conformal boundary which is transverse to the $t$ axis, pictured as extending through the bulk from one pole to the other. Then $t = -\infty$ is one half of the boundary, and $t = +\infty$ is the other $-$ and we see at once that, despite the appearance of (32), the boundary is connected. The effective connectedness of the boundary of the Lorentzian version of deSitter space is reflected in the actual connectedness of the boundary of its Euclidean version. (The reader will have noted the close analogy with our discussion of the Karch-Randall theory, and indeed the latter may have much to teach us about the dS/CFT correspondence.)
\ss
Of course we would like this happy agreement to persist when we turn to {\it asymptotically} deSitter spacetimes. Now, as always, the statement that a spacetime is ``asymptotically" the same as some given spacetime has two aspects, geometric and physical. Geometrically it is straightforward to formulate a definition of the asymptotic deSitter condition, but this has to be supplemented by a physical condition on the rate at which the metric in question approaches the deSitter metric. Let us consider whether the Smash-free Phantom metrics $g(\gamma, A)$ satisfy all of these conditions.
\ss
That the Lorentzian geometry of $Ph(\gamma, A)$ approaches $dS_4$ geometry towards temporal infinity has already been discussed, so we concentrate on the (negative) Euclidean version. Recall that $dS_4$ itself becomes $-H^4$ under negative Euclideanization (that is, triple analytic continuation), so we need a definition of what it means for a negative Riemannian manifold to be {\it asymptotically negative hyperbolic}. Briefly, the definition [35] is as follows. Let $W^{n+1}$ be a non-compact $(n+1)$-dimensional manifold which can be regarded as the interior of a compact, connected manifold-with-boundary ${\overline W}^{n+1}$, and let $N^n$ be the boundary (which need not be connected).  Let $g^{-W}$ be a smooth negative metric on $W^{n+1}$ such that there exists a smooth function $F$ on ${\overline W}^{n+1}$ with the following properties:
\ss
\item{(i)}		$F(x) = 0$ if and only if $x \in N^n$;

\item{(ii)}	$dF(x) \ne 0$ for all $x \in N^n$;

\item{(iii)}	$F^2g^{-W}$ extends continuously to a negative metric on 
			${\overline W}^{n+1}$;

\item{(iv)}	If $|dF|_F$ is the norm of $dF$ with respect to the extended 			negative metric, then $|dF|_F$, evaluated on $N^n$, must not 			depend on position there.

\ss
\nt The first three conditions ensure that the boundary is infinitely far from any point in the interior. So $-W^{n+1}$ (that is, $W^{n+1}$ endowed with $g^{-W}$) may be called an {\it asymptotically negative hyperbolic space}; for all sectional curvatures along geodesics ``tending to infinity'' approach a common {\it positive} constant, the asymptotic sectional curvature, which is equal to the square of $|dF|_F$, evaluated on $N^n$. (Notice that it follows from this last fact that a non-compact flat Euclidean space cannot have a conformal compactification. For the vanishing of the asymptotic sectional curvature would imply that $|dF|_F = 0$, and in (positive or negative) Euclidean geometry that would contradict part (ii) of the definition.)
\ss
Now the Euclidean version of $Ph(\gamma, A)$ is of course obtained by complexifying the spatial coordinates in equation (22), which gives us simply
$$g^{Euc}(\gamma ,A) = -dt \otimes dt - A^2 {\cosh}^{\left({4\over \gamma}\right)}
	\left( {\gamma t \over 2L}\right) (d\theta_1 \otimes d\theta_1 + 
d\theta_2 \otimes d\theta_2 + d\theta_3 \otimes d\theta_3). \eqno{(33)}$$
It is straightforward to verify that this metric, with $t$ allowed to run from $-\infty$ to $+\infty$, is indeed asymptotically negative hyperbolic; the asymptotic sectional curvature, computed by evaluating the square of $|dF|_F$, is just $+1/L^2$, for all $\gamma$ and $A$. However, it is essential here that $\theta_1$, $\theta_2$, and $\theta_3$ should retain their angular interpretation; otherwise the slices would be non-compact flat spaces, and, as we just saw, these do not have a conformal compactification in the Euclidean case. That is, the slices continue to be tori. 
\ss
Next, we turn to the physical aspect of the definition of ``asymptotically deSitter". This was considered in [15] and [29], where it was argued that the correct definition must be given in terms of the deSitter version [25] of the Brown-York stress-tensor, as applied to the AdS/CFT correspondence [38]: the eigenvalues must tend to finite values at infinity. Hence we must verify this for $Ph(\gamma, A)$. 
\ss
According to [25], the Brown-York tensor at future infinity of a 
four-dimensional spacetime with asymptotic Ricci tensor eigenvalues $3/L^2$ (which will be the case for any spacetime with a negative Euclidean version which is asymptotically negative hyperbolic with asymptotic sectional curvature equal to $1/L^2$) is determined by evaluating the tensor
$$T^\mu_\nu = {1\over 8\pi} [K^\mu_\nu - (K + {2\over L})\delta^\mu_\nu - LG^\mu_\nu] \eqno{(34)}$$
on a spacelike hypersurface and then taking time to infinity. Here the indices run from 1 to 3,
$K^\mu_\nu$ is the extrinsic curvature of the hypersurface, $K$ is its trace, and $G^\mu_\nu$ is the Einstein tensor of the hypersurface. The analogous formula (without the Einstein tensor term) yields, for $dS_3$ with the spherical slicing [25],    
$$T^\mu_\nu (dS_3; \hbox{spherical slicing}) = -{1\over 8\pi} 
{e^{-t/L}\over L{\cosh}\left({t\over L}\right)} \delta^\mu_\nu . \eqno{(35)}$$
The central charge of the corresponding CFT should be given [39] by the limiting value of the quantity $-24\pi (T/R)$, where $T$ is the trace of the stress tensor and $R$ is the scalar curvature of the hypersurface. We have
$$-24\pi (T/R) = {3L \over 2} \left(1 + e^{-{2t\over L}}\right), \eqno{(36)}$$
which indeed gives the correct [15] value, $3L/2$, when $t$ is taken to infinity [40]. Notice that the final value is approached from above; the Brown-York expression {\it decreases} to its asymptotic value as time goes on. (We would like to recommend that, in using the Brown-York tensor in this way, one should check not just the limiting values but also the way in which they are approached. It seems quite possible that these additional data could be useful.) In any case, the fact that this calculation gives a reasonable answer does support the contention that the Brown-York tensor is the correct measure of the stress-energy of the CFT at infinity; in which case, the requirement that it should be finite is a reasonable condition to attach to the definition of an asymptotically deSitter spacetime.
\ss
For our purposes, however, we wish the spacelike hypersurfaces to be flat. With the flat slicing, we obtain for $dS_4$ the result [25]
$$T^\mu_\nu (dS_4; \hbox{flat slicing}) = 0.  \eqno{(37)}$$
That is, the Brown-York tensor vanishes {\it even before} we take the limit. For the metric of $Ph(\gamma, A)$ we obtain, by contrast,
$$T^\mu_\nu (Ph(\gamma ,A)) = {-1\over 4\pi L} \left( 1 - {\tanh}\left({\gamma t \over 2L}\right)\right) \delta^\mu_\nu .     \eqno{(38)}$$
Thus we find that the eigenvalues of the Brown-York tensor are all {\it negative} at any finite time, and they all tend to a finite value, namely zero, as $t$ tends to infinity; this for all $\gamma$ and $A$. We conclude that $Ph(\gamma, A)$ is indeed asymptotically deSitter, in every sense, for all $\gamma$ and $A$. Hence, the dS/CFT correspondence should make sense for these spacetimes. (Before moving on, we note that the ``mass conjecture" of [25] suggests that the ``mass" of any non-singular asymptotically deSitter spacetime (defined in terms of the limit of the Brown-York tensor) should not exceed that of deSitter space, namely zero in this case. The fact that (38) tends to zero is therefore consistent with the conjecture, since the Smash-free cosmologies are completely non-singular; and perhaps the fact that zero is approached from below might also be considered appropriate from that point of view. 
\ss
Now we saw in Section 2 that the Penrose diagram of $Ph(\gamma, A)$ is of finite extent in conformal time, with null curves connecting past infinity to future infinity. (If, as we are assuming, $\gamma$ is small, so that the diagram is ``tall and thin", then a ray of light will be able to circumnavigate the (toral) universe several times.) As in the case of deSitter space itself, therefore, the conformal boundary of $Ph(\gamma, A)$ is effectively connected. But now, when we examine equation (33) giving the Euclidean version, we see that its conformal boundary is {\it not} connected: there is a torus at past infinity, and another at future infinity. This is of course in sharp contrast to the case of deSitter space, the Euclidean version of which has a connected conformal boundary. This is an interesting situation, mathematically and physically.
\ss
Taking first a mathematical standpoint, it is of course well known that Witten and Yau [19] attempted to prove that such manifolds can be ruled out by imposing well-motivated conditions on the boundary geometry. Their results have stimulated the growth of a new field of geometric analysis [22][23][41]. Most work on this subject assumes that the manifold is Einstein, but Cai and Galloway [22] extended the main result to non-Einstein manifolds, while also significantly weakening the condition on the boundary. Translated into the negative Riemannian language we need here, their result may be stated as follows. Let $-W^{n+1}$ be a connected complete (n+1)-dimensional asymptotically negative hyperbolic manifold with asymptotic sectional curvature $+1/L^2$.  The eigenvalue functions of the $(1,1)$ version of the Ricci tensor, $\lambda_\beta$, are then asymptotic to $n/L^2$.  Suppose that, with a canonical choice of defining function $F$,
$$F^{-4}\left(\lambda_\beta - {n\over L^2}\right) \to 0 \eqno{(39)}$$
uniformly as the boundary is approached.  Suppose too that
$$\lambda_\beta \le {n \over L^2} \eqno{(40)}$$
for all $\beta$ and that the negative conformal structure induced on any connected component of the conformal boundary $-N^n$ is represented by a (negative) metric of non-positive scalar curvature.  Then $-N^n$ is connected. Notice that the condition of {\it completeness} is essential here, so this theorem cannot constrain the incomplete Euclidean spaces used in [21] to construct a Hartle-Hawking state. That is, Maldacena's space evades the theorem by not obeying one of its assumptions. We shall now see that $Ph(\gamma, A)$ evades it by failing to obey another.
\ss
Now $Ph(\gamma, A)$ is certainly connected, complete, and asymptotically negative hyperbolic with asymptotic sectional curvature $+1/L^2$. Furthermore, the (negative) conformal structure induced on the boundary components is obviously represented by a metric of non-positive scalar curvature $-$ the flat metric on the torus. The Ricci tensor of the Euclidean metric (33) is given by, in an obvious notation,
$$\hbox{Ric} (g^{\rm Euc}(\gamma ,A))^t_t = {3\over L^2} - {3\over L^2} \left( 1 - {\gamma \over 2}\right) a^{-\gamma}    \eqno{(41)}$$
and
$$\hbox{Ric} (g^{\rm Euc}(\gamma ,A))^\mu_\nu = \left({3\over L^2} - {3\over L^2} \left( 1 - {\gamma \over 6}\right) a^{-\gamma}\right)   \delta^\mu_\nu,   \eqno{(42)}$$
where $\mu$ and $\nu$ run from 1 to 3. Clearly condition (40) will be satisfied, with $n = 3$, provided that condition (11) is slightly strengthened to 
$$ 0 < \gamma \leq 2,   \eqno{(43)}$$
which we shall assume henceforth, since we want $\gamma$ to be very small in any case. (Notice that, from equation (27), this is just the condition that the acceleration of the Universe should always be slower than that of the corresponding deSitter spacetime). So every condition of the Witten-Yau-Cai-Galloway theorem is satisfied $-$ except (39). Clearly $a^{-1}$ plays the role of $F$ here, and (41),(42), and (43) conspire to ensure that (39) is {\it not} satisfied. This is how the Euclidean version of $Ph(\gamma, A)$ is able to have a disconnected conformal boundary: its Ricci eigenvalues do not decay to the deSitter values sufficiently rapidly. However, it is not easy to see how to justify any requirement that they should do so; we have already established that the Lorentzian version satisfies the physical conditions for an asymptotically deSitter spacetime.
\ss
According to our earlier discussion, the disconnectedness of the conformal boundary of the Euclidean version of $Ph(\gamma, A)$ has two possible interpretations: either the Lorentzian version is in some way physically unacceptable, or the two boundary components are inhabited by independent but quantum-mechanically entangled CFTs. We now consider these possibilities in turn.
\bs
\nt{\bf 4.\ \ TIME AS AN INVERSE RENORMALIZATION GROUP FLOW}
\ms
It has often been observed that the present cosmic acceleration is very much like a slower version of cosmic inflation. In the context of the dS/CFT correspondence, it is natural to ask what this observation means from the CFT point of view. This question was raised in both [24] and [25], and the conclusion was as follows. The Bekenstein-Hawking formula, applied to the deSitter horizon, indicates that since the effective cosmological constant was larger in the (inflationary) past, the number of degrees of freedom was small; so from the boundary CFT point of view, the remote past corresponds to an infrared fixed point. Similarly, the much slower cosmic acceleration at the present time corresponds to a much larger number of degrees of freedom, and the remote future presumably corresponds to an ultraviolet fixed point. Since a renormalization group flow is from the UV to the IR, one concludes that {\it the passage of time is an inverse renormalization group flow}. This is a truly deep application of the dS/CFT correspondence, for it means that the passing of time is a {\it derived} concept: ultimately it arises from the properties of a CFT {\it which is itself timeless}. It follows that if the correspondence can be applied to a given spacetime, then the direction of time flow is not freely prescribable, as it is in classical cosmology. The direction of ``time" flow is necessarily from the IR to the UV, that is, by the Bekenstein-Hawking formula, from large values of $|H|$, where $H$ is the Hubble parameter, to small. (See [42] for a concrete model of this
scenario.)
\ss
There are however two subtleties here. First, as always in general relativity, when speaking of ``time", we must ask, ``which time?" As is emphasized in [24], in order to use this argument we must use a definition of bulk time such that evolution along this time induces a conformal transformation of the boundary, as does deSitter time in the case of the metric given by equation (23). The above prescription therefore applies most clearly to spacetimes which closely resemble (23) near a component of the boundary; such spacetimes have flat spatial sections. (But see [17].) We can in such cases use the prescription to determine the direction of time flow {\it near} the boundary. The idea then is to use the irreversibility of the renormalization group flow to extrapolate the result into the bulk, thereby fixing the direction of time flow far from the boundary. 
\ss
The second objection to this approach, also raised in [24], is that we are assuming that the boundary CFT behaves in the conventional way. However, the computation of conformal weights corresponding to a scalar field propagating in $dS_4$ indicates [15] that the boundary CFT will only be unitary if the mass of the scalar is no greater than $3/2L$. Hence we cannot always be confident that the CFT behaves in a reasonable way unless any scalar matter we introduce has a very small mass, with a Compton wavelength about the size of the horizon. 
\ss
In the case of $Ph(\gamma, A)$, we saw in Section 2 that $g(\gamma, A)$ does indeed resemble (23) at large values of $t$. Furthermore, the ``matter" we are introducing in order to violate the DEC may be a kind of negative-energy quintessence, which is in fact just an extremely light scalar field with a Compton wavelength of order the size of the horizon. It {\it may} therefore be safe to assume that the direction of time flow in $Ph(\gamma, A)$ can be determined by examining the behaviour of $|H|$ at large values of $t$. From equation (21), we have 
$$|H(\gamma)| = {1 \over L}|{\tanh}({\gamma t \over 2L})|.   \eqno{(44)}$$ 
This makes sense for negative $t$, since $|H|$ clearly decreases as $t$ increases, as it should according to the above argument. But beyond $t = 0,$ $|H|$ begins to {\it increase}, which simply does not make sense according to the holographic interpretation $-$ it contradicts the irreversibility of the renormalization group flow of the CFT. Thus, the ``upper half" of $Ph(\gamma, A)$ cannot be interpreted as an expanding universe. It can be given a physical meaning only by detaching it from the lower half and by replacing $t$ by $-t$. But then we just obtain another, independent copy of the lower half. The latter represents a contracting universe whose evolution comes to a halt at $t = 0$ (because $|H|$ has its minimum value there, so the renormalization group flow stops.)
\ss
From the viewpoint of a conventional CFT, then, $Ph(\gamma, A)$ is obtained by artificially conjoining two copies of a contracting cosmology along the surface where $t = 0$. The bulk seemed to be connected when we adopted the classical interpretation of time flow; but now that this has been corrected by the dS/CFT correspondence, we see that it is in fact {\it effectively disconnected}, just as the Euclidean version predicts. 
\ss
{\it If} this is the correct interpretation, then our Phantom cosmology must be rejected, for it predicts that the Universe must necessarily contract. The alternative interpretation is that the situation here is a kind of ``ninety-degree rotation" of the one considered by Maldacena in [21]: we really do have two independent CFTs on the two boundary components (separated now by
time instead of space $-$ hence the ``ninety-degree rotation" picture), and in both cases this is acceptable because the CFTs are entangled. This possibility has come to seem very much more plausible with the appearance of [43], where this very possibility is considered. It is found there that, in order to construct a dS/CFT correspondence consistent with this entanglement, one must introduce CFTs with very unfamiliar properties: for example, these new CFTs even require a re-formulation of the way conformal symmetry acts, involving a ``Euclidean Virasoro algebra". Such a radical approach can be expected to yield some surprises; it could well lead to unusual behaviour of the renormalization group flow. Hence our argument above, based on assuming that the renormalization group flow of these CFTs behaves in a familiar way, becomes very questionable. It is conceivable, for example, that the relation between time direction and CFT flow could differ before and after the ``bounce"; one would have to give an explicit example of a flow in an entangled state in a product of such special CFTs to determine this. (The connectedness of the Euclidean boundary for deSitter space itself would then indicate that time would not flow in pure deSitter space $-$ not an entirely unreasonable result, in view of the fact that deSitter space contains no matter other than the cosmological constant dark energy, which, with its rigidly constant density and pressure, could not register any time flow.)
While it is premature to try to guess whether this possibility can rescue the Smash-free Phantom cosmology, what can be said is that this is definitely the most promising way to attempt such a rescue. The alternative is a Big Smash: a grim prospect from any point of view.
\bs
\nt{\bf 5.\ \ CONCLUSION}
\ms
The fundamental observation of this work is that Anti-deSitter space violates the Dominant Energy Condition, and deSitter space is on the brink of violating it. Asymptotically deSitter spacetimes therefore fall into two ``equally large" classes: those which do not violate the DEC, and those which do. We could choose to ignore the latter class, but that would be perilous both observationally and theoretically. If we choose not to ignore it, the most natural way to model its members in the context of string theory is to introduce DEC-violating ``matter" into a spacetime with a positive cosmological constant. The resulting ``Smash-free" Phantom cosmologies were constructed by assuming, motivated by the observational evidence, that the spatial sections are flat. They appear to be reasonable, but the Euclidean formulation suggests that all is not well with them; and the methods of [24] and [25] seem to confirm this. The Einstein equation for FRW cosmologies with flat spatial sections (equation (18)) indicates that this problem afflicts all spatially flat Phantom cosmologies.
\ss
Negatively curved spatial sections are also forbidden by the dS/CFT correspondence. For such a spacetime would induce at infinity a conformal structure represented by a metric of negative scalar curvature (positive in the ``negative Euclidean" formulation), and it is well known that this would cause instability both on the boundary and in the bulk [18]. Positively curved spatial sections lead to an Einstein equation of the form
$${{3{\mathop{a}\limits^\cdot}^2} \over a^2} = 8\pi [\rho - {{3/8\pi L^2}\over a^2}]. \eqno{(45)}$$
Thus we see that deSitter density in this case is already achieved at $t = 0$; hence it cannot {\it increase} to that value as equation (7) requires. We conclude that {\it if the boundary CFTs behave in the conventional way,} then FRW models of Smash-free Phantom cosmology are not viable. 
\ss
With the appearance of [43], however, we now have another possibility: the Smash-free cosmology could be rescued by means of a deeper understanding of the boundary CFTs, which appear to differ very radically from those to which the AdS/CFT correspondence has accustomed us. Whether this rescue can actually be effected will depend on whether we can reliably calculate a flow in an entangled state in a product of these special CFTs.
\ss
Non-singular ``bouncing" cosmologies which violate the DEC have recently attracted much attention [26], and a glance at the analysis of [4] shows that there is a very real possibility that the cosmological equation-of-state parameter satisfies $w < -1$. If that should prove to be the case, we shall be faced with one of two very remarkable situations. The first is the prospect of a Big Smash, an all-encompassing singularity which will actually occur in the finite future, and which (since it involves universal expansion instead of compression) will probably not be amenable to a quantum-mechanical analysis or resolution. The second is the possibility that we reside in a ``bouncing" universe completely controlled by a pair of quantum-entangled CFTs, of a radically new kind, residing in the infinite future and past. The second scenario, with its singularity-free spacetime and its extremely deep relation between cosmic evolution and quantum mechanics, is doubtless preferable, but much remains to be done to see whether it can work. In exploring it, we shall need a $``w < -1"$ analogue of deSitter space; and the metrics $g(\gamma, A)$ given here are natural candidates for that role.

\bs
\nt{\bf REFERENCES}
\ms
\item{[1]} J. L. Sievers, J. R. Bond, J. K. Cartwright, C. R. Contaldi, B. S. Mason, S. T. Myers, S. Padin, T. J. Pearson, U.-L. Pen, D. Pogosyan, S. Prunet, A. C. S. Readhead, M. C. Shepherd, P. S. Udomprasert, L. Bronfman, W. L. Holzapfel, J. May, Cosmological Parameters from Cosmic Background Imager Observations and Comparisons with BOOMERANG, DASI, and MAXIMA,
astro-ph/0205387.
\item{[2]} C. Barcelo, Matt Visser, Twilight for the energy conditions? gr-qc/0205066.
\item{[3]} R.R. Caldwell, A Phantom Menace?, astro-ph/9908168.
\item{[4]} S. Hannestad, E. Mortsell, Probing the dark side: Constraints on the dark energy equation of state from CMB, large scale structure and Type Ia supernovae, astro-ph/0205096. 
\item{[5]} B. Boisseau, G. Esposito-Farese, D. Polarski, A.A. Starobinsky, Reconstruction of a scalar-tensor theory of gravity in an accelerating universe, Phys.Rev.Lett. 85 (2000) 2236,   gr-qc/0001066.
\item{[6]} V. Faraoni, Superquintessence, Int.J.Mod.Phys. D11 (2002) 471, astro-ph/0110067. 
\item{[7]} V. K. Onemli, R. P. Woodard, Super-Acceleration from Massless, Minimally Coupled $\phi^4$, gr-qc/0204065 
\item{[8]} S. Kachru, L. McAllister, Bouncing Brane Cosmologies from Warped String Compactifications, hep-th/0205209.
\item{[9]} V. Sahni, Y. Shtanov, Braneworld models of dark energy, astro-ph/0202346. 
\item{[10]} S. W. Hawking, G. F. R. Ellis, The Large Scale Structure of Space-Time, 		Cambridge University Press, 1973.
\item{[11]} B. Carter, Energy dominance and the Hawking Ellis vacuum conservation theorem,
gr-qc/0205010. 
\item{[12]} A.A. Starobinsky, Future and Origin of our Universe: Modern View, Grav.Cosmol. 6 (2000) 157, astro-ph/9912054. 
\item{[13]} C.M. Hull, Timelike T-Duality, de Sitter Space, Large $N$ Gauge Theories and Topological Field Theory, JHEP 9807 (1998) 021, hep-th/9806146.
\item{[14]} E. Witten, Quantum Gravity In De Sitter Space, hep-th/0106109.
\item{[15]} A. Strominger, The dS/CFT Correspondence, JHEP 0110 (2001) 034, hep-th/0106099.
\item{[16]} A.J.M. Medved, A Holographic Interpretation of Asymptotically de Sitter Spacetimes,
Class.Quant.Grav. 19 (2002) 2883, hep-th/0112226.
\item{[17]} F.Leblond, D.Marolf, R. C. Myers, Tall tales from de Sitter space I: Renormalization group flows, JHEP 0206 (2002) 052, arXiv:hep-th/0202094. 
\item{[18]} E.Witten, http://online.itp.ucsb.edu/online/susy\_c99/witten/
\item{[19]} E. Witten, S. T. Yau, Connectedness of the Boundary in the $AdS/CFT$ 		Correspondence, Adv.Theor. Math. Phys. 3 (1999) 1635, hep-th/9910245.
\item{[20]} V. Balasubramanian, P. Kraus, A. Lawrence, S. Trivedi, Holographic Probes of Anti-de Sitter Spacetimes, Phys.Rev. D59 (1999) 104021, hep-th/9808017.
\item{[21]} J.M. Maldacena, Eternal Black Holes in AdS, hep-th/0106112.
\item{[22]} M. Cai, G. J. Galloway, Boundaries of Zero Scalar Curvature in the $AdS/CFT$ Correspondence, Adv.Theor.Math.Phys. 3 (1999) 1769, hep-th/0003046.
\item{[23]} M. T. Anderson, Boundary regularity, uniqueness and non-uniqueness for AH Einstein metrics on 4-manifolds, math.DG/0104171. 
\item{[24]} A. Strominger, Inflation and the dS/CFT Correspondence, JHEP 0111 (2001) 049, hep-th/0110087.
\item{[25]} V. Balasubramanian, J. de Boer, D. Minic, Mass, Entropy and Holography in Asymptotically de Sitter Spaces, Phys.Rev. D65 (2002) 123508, hep-th/0110108.
\item{[26]} S. Tsujikawa, R. Brandenberger, F. Finelli, On the Construction of Nonsingular Pre-Big-Bang and Ekpyrotic Cosmologies and the Resulting Density Perturbations, hep-th/0207228 
\item{[27]} J. Levin, Topology and the Cosmic Microwave Background, Phys.Rept. 365 (2002) 251, gr-qc/0108043.
\item{[28]} G. Galloway, K. Schleich, D. Witt, E. Woolgar, The $AdS/CFT$ 			Correspondence Conjecture and Topological Censorship, Phys.Lett. 			B505 (2001) 255, hep-th/9912119.
\item{[29]} M. Spradlin, A. Strominger, A. Volovich, Les Houches Lectures on De Sitter Space, hep-th/0110007.
\item{[30]} A. Karch, L. Randall, Locally Localized Gravity, JHEP 0105 (2001) 008, hep-th/0011156.
\item{[31]} A. Karch, L. Randall, Open and Closed String Interpretation of SUSY CFT's on Branes with Boundaries, JHEP 0106 (2001) 063, hep-th/0105132.
\item{[32]} M. Porrati, Mass and Gauge Invariance IV (Holography for the Karch-Randall Model), 
Phys.Rev. D65 (2002) 044015, hep-th/0109017.
\item{[33]} M.J. Duff, J.T. Liu, Complementarity of the Maldacena and Randall-Sundrum Pictures, Phys.Rev.Lett. 85 (2000) 2052-2055; Class.Quant.Grav. 18 (2001) 3207-3214, hep-th/0003237.
\item{[34]} E. Witten, Anti-deSitter Space and Holography, Adv. Theor. Math. Phys. 2 (1998) 253, hep-th/9802150.
\item{[35]} B. McInnes, Exploring the Similarities of the dS/CFT and AdS/CFT Correspondences, Nucl.Phys. B627 (2002) 311, hep-th/0110062.
\item{[36]} D. Birmingham, M. Rinaldi, Bubbles in Anti-de Sitter Space, hep-th/0205246. 
\item{[37]} A. J. Tolley, N. Turok,  Quantization of the massless minimally coupled 		scalar field and the $dS/CFT$ correspondence,
	     hep-th/0108119.
\item{[38]} V. Balasubramanian, P. Kraus, A Stress Tensor for Anti-de Sitter Gravity, Commun.Math.Phys. 208 (1999) 413, hep-th/9902121.
\item{[39]} S. Nojiri, S.D. Odintsov, Conformal anomaly from dS/CFT correspondence, Phys.Lett. B519 (2001) 145, hep-th/0106191.
\item{[40]} D. Klemm, Some Aspects of the de Sitter/CFT Correspondence, Nucl.Phys. B625 (2002) 295, hep-th/0106247.
\item{[41]} N.C. Leung, T. Y.-H. Wan, Harmonic maps and the topology of conformally compact Einstein manifolds, math.DG/0112023.
\item{[42]} A. C. Petkou, G. Siopsis, dS/CFT correspondence on a brane, JHEP 0202 (2002) 045, hep-th/0111085.
\item{[43]} V. Balasubramanian, J. de Boer, D. Minic, Exploring de Sitter Space and Holography,
hep-th/0207245 
\bye